\documentclass[aps]{revtex4}
\usepackage{graphicx}
\usepackage[colorlinks=true,urlcolor=blue,linkcolor=blue,citecolor=green]{hyperref}
\newcommand{\beq}{\begin{equation}}
\newcommand{\eeq}{\end{equation}}
\newcommand{\inia}{\begin{eqnarray}}
\newcommand{\fina}{\end{eqnarray}}

\begin{document}

\draft
\title{\bf Gauge field localization on brane worlds }
\author{Rommel Guerrero $^1$, Alejandra Melfo $^{2}$, Nelson Pantoja $^2$
and R. Omar Rodriguez $^1 $}
\address{ $^1$ Departamento de F\'isica,
Universidad Centroccidental Lisandro Alvarado, 400 Barquisimeto,
Venezuela}
\address{ $^2$ Centro de F\'isica Fundamental, Universidad de Los Andes,
M\'erida, Venezuela}

\begin{abstract}

We consider the effects of spacetime curvature and brane thickness on the localization of gauge fields on a brane {\it via}   kinetic terms induced by localized fermions. 
  We find that in a warped geometry with and infinitely thin brane, both the infrared and the ultraviolet behavior of the electromagnetic propagator are affected, providing a more stringent bound on the brane's tension than that coming from the  requirement of four-dimensional gravity on the brane.  On the other hand, for a thick wall  in a  flat spacetime, where the fermions are localized by means of a Yukawa coupling, we find that  4-dimensional electromagnetism is recovered  in a region bounded from above by the same critical distance appearing in the thin case, but also from below by a new scale related to the brane's thickness and the electromagnetic couplings.

\vspace{0.5 cm} 
\end{abstract}

\maketitle

\vspace{0.3cm}

\section{Introduction}

If our Universe is to  be realized on a brane embedded in a higher dimensional spacetime, the Standard Model fields should be confined to the brane, as should four dimensional gravity. In the Randall-Sundrum 2 (RS-2) scenario \cite{Randall:1999vf}, the gravitons exhibit a normalizable zero mode which reproduces four dimensional gravity on the brane, plus a continuous spectrum which gives corrections to this four dimensional behaviour. Localization of fermions and vector fields in this scenario requires, however, non-gravitational mechanisms \cite{Bajc:1999mh}. In domain wall realizations of the RS-2 scenario, after introducing Yukawa interactions of the fermions with the scalar field from which the wall is made of \cite{Rubakov:1983bb}, fermions of one chirality can be confined \cite{Ringeval:2001cq, Koley:2004at, Melfo:2006hh}. Naturally, additional efforts have been made to find mechanisms which confine also gauge fields to the wall. 

In absence of gravity, there are several proposals to achieve this. For example, massless gauge particles can be localized on a brane under the assumption that is possible to have a gauge theory that is confining in the bulk and with no confinement on the brane \cite{Dvali:1996xe}. Another interesting approach is the proposal of Dvali, Gabadadze and Shifman (DGS) \cite{Dvali:2000rx}, where the addition of a brane localized kinetic term to the conventional five dimensional gauge field action results in a gauge field propagator that provides, below a critical distance, the correct four-dimensional electromagnetic potential (see Ref.\cite{Dvali:2000hr} for the gravitational analogue). Remarkably, both mechanisms preserve the charge universality condition \cite{Dubovsky:2001pe}. 

The phenomelogical consequences of  brane localized kinetic terms for gauge fields which can propagate in the bulk of the two branes Randall-Sundrum (RS-1) scenario \cite{Randall:1999ee}, where the extra dimension is compactified on $S_1/Z_2$, have been examined in Refs.\cite{Davoudiasl:2002ua,Carena:2002dz}. On the other hand, it has been shown that the introduction of additional bulk fields may lead, through its couplings, to localized gauge fields in the RS-2 scenario \cite{Kehagias:2000au,Oda:2001ux,Ghoroku:2001zu}. 

In this paper we extend to curved spacetime and thick walls the DGS flat model of Ref.\cite{Dvali:2000rx}. In Section II we provide the general framework for computing the gauge field propagator in a warped spacetime generated by a thick domain wall, including in the effective action a four-dimensional induced gauge kinetic term generated by matter fields confined on the wall. The DGS result is shown to be recovered in the thin-wall, flat space limit. 
 
In section III we then consider the gauge field propagator on the RS-2 brane world, for matter that is prescribed on the infinitely thin wall or brane by a Dirac's delta distribution. We found that the inclusion of gravitation modifies both the ultraviolet and infrared behavior of the effective four dimensional gauge field propagator. In particular, the critical distance below which four dimensional electromagnetism is recovered depends exponentially on the cosmological constant and the electromagnetic 4 and 5-dimensional couplings. As could be expected there is also a minimal scale, the RS critical distance, below which again 5 dimensional effects are found in the propagator.  
 
Next, in section IV, the brane is considered as a thick domain wall generated by a scalar field with a suitable symmetry breaking potential and the fermions coupled to the wall by a Yukawa term (see for example Refs.\cite{Ringeval:2001cq, Koley:2004at, Melfo:2006hh,Liu:2009dw}). The 4-dimensional propagator can be found exactly in the flat case for particular values of the Yukawa coupling.  We demonstrate the existence of a region, bounded by two critical scales, where electromagnetism has a four-dimensional behavior. The upper bound corresponds to the critical radius given by the DGS model; while the lower bound is determined by the bulk gauge coupling constant and coincides with the one estimated by Dubovsky and Rubakov  \cite{Dubovsky:2001pe}. We argue that this imposes stringent bounds on the wall's thickness which seem to invalidate the DGS mechanism for this particular case, although in the general scenario it depends non trivially on the Yukawa coupling and the parameters of the scalar field potential.

\section{Gauge fields on a gravitating thick brane}\label{DGS}

In the DGS proposal \cite{Dvali:2000rx}, matter is supposed to be confined to a four-dimensional slice of 
Minkowski spacetime.  As consequence of quantum fluctuations of the matter fields, the bulk gauge field propagator gets corrections on the brane world volume, resulting in an ordinary four-dimensional Coulomb potential on the brane for all distances below a critical radius. The critical radius  is inversely proportional to the square of the bulk gauge coupling constant, which can then be adjusted to account for (non) observation of deviation of electromagnetism at large scales in our Universe. 

We wish  to extend the DGS brane-induced localization mechanism of vector fields in flat space \cite{Dvali:2000rx} to the case of a brane embedded in a curved spacetime with a warped metric of the form  
\begin{equation}
ds^2= e^{2 a(y)}\ \eta_{\mu\nu}dx^{\mu}dx^{\nu}+ dy^2, \label{metric}
\end{equation}
where $\eta_{\mu\nu}$ is the four-dimensional Minkowski metric. So we will consider the five-dimensional effective action
\begin{equation}\label{effective}
S=\int dx^4\ dy \sqrt{g}\left[-\frac{1}{4{\cal Q}^2}F_{ab}F^{ab}-\frac{1}{4 e^2} k^2(y) F_{\mu\nu}F^{\mu\nu}-J^c(x,y) A_c\right],
\end{equation}
where ${\cal Q}$ and $e$ are the five and four-dimensional gauge couplings, respectively, as our starting point to study the localization of gauge fields on thick branes embedded in a curved spacetime (latin and greek indices correspond to five and four-dimensions respectively). For $g_{ab}=\eta_{ab}$, with $\eta_{ab}$ the 5-dimensional Minkowski metric and $k^2(y)=\delta(y)$, we have the effective action proposed in \cite{Dvali:2000rx} which lends to quasi-localization of gauge fields on a thin wall at the tree-level. However, it should be noted that the physical significance of the case $k^2(y)=\delta(y)$ is no obvious in a curved space. In particular, for BPS branes embedded in a curved spacetime, it has been shown that chiral fermions coupled to the brane's scalar field through a Yukawa term, require Yukawa couplings to localize them that diverge in the thin-wall limit \cite{Melfo:2006hh}. Nevertheless, as we will see, $k^2(y)$ is not the profile of the energy density of a thick wall in the gravitating case and may have an approximated $\delta$-like profile even for finite-thickness domain walls. 

Next, in order to determine the tree-level propagator $G_{ab}$ for the gauge field, we introduce in the effective action the gauge-fixing term $-\frac{1}{2\xi}(n^bA_b)^2$ with $n^b=m\delta_y^b$ and $\xi\rightarrow 0$ corresponding to the unitary gauge $A_y=0$. We  find for the $(y,b)$ components
\begin{equation}
e^{-2a(y)}\eta^{\alpha\beta}\partial_{\beta}(\partial_{\alpha}G_{yb}-\partial_yG_{\alpha b}) - \frac{{\cal Q}^2}{\xi}mG_{yb}=g_{yb}{\cal Q}^2(\sqrt{g})^{-1}\,\delta^4(x-x')\delta(y-y').
\end{equation}
Fourier-transforming the brane coordinates this equation gives $\tilde{G}_{y\alpha} \to 0$ for $\xi\rightarrow 0$ and we thus find that $\tilde{G}_{\alpha\beta}$ in the unitary gauge satisfies
\begin{equation}
(-\bar{p}^2\eta^{\sigma\beta} + \bar{p}^{\sigma}\bar{p}^{\beta})\tilde{G}_{\beta\gamma} + \eta^{\sigma\beta}\partial_y(e^{2a(y)}\partial_y\tilde{G}_{\beta\gamma}) + k^2(y)\frac{{\cal{Q}}^2}{e^2}(-\bar{p}^2\eta^{\sigma\beta} + \bar{p}^{\sigma}\bar{p}^{\beta})\tilde{G}_{\beta\gamma}={\cal{Q}}^2\delta^{\sigma}_{\gamma}\delta(y-y'), 
\end{equation}
where $\bar{p}^{\alpha}\equiv\eta^{\alpha\beta}p_{\beta}$ and $\bar{p}^2\equiv\eta^{\alpha\beta}p_{\alpha}p_{\beta}$.

Now, writing $\tilde{G}_{\beta\gamma}$ in the form
\begin{equation}\label{four-propagator}
\tilde{G}_{\beta\gamma}=\left(\eta_{\beta\gamma} - \frac{p_{\beta}p_{\gamma}}{\bar{p}^2}\right)G_1 + \frac{p_{\beta}p_{\gamma}}{\bar{p}^2}G_2
\end{equation}
it can be seen that $G_1$ and $G_2$ satisfy
\begin{equation}
\partial_y\left(e^{2a(y)}\partial_yG_1\right) -\left(1 + k^2(y)\frac{{\cal{Q}}^2}{e^2}\right)\bar{p}^2G_1= {\cal{Q}}^2 \delta(y-y')
\label{g1}
\end{equation}
and
\begin{equation}
\partial_y\left(e^{2a(y)}\partial_y G_2\right) = {\cal{Q}}^2 \delta(y-y')
\end{equation}
respectively. Note that $G_2$ is independent of the momentum.

Next, in the background (\ref{metric}), let us consider the generating functional
\begin{equation}\label{generating}
W[J]=W[0]\exp\left\{-\frac{i}{2}\int d^4x \,dy\sqrt{g(y)}\int d^4x' dy' \sqrt{g(y')}J^a(x,y)G_{ab}(x-x',y,y')J^b(x',y')\right\}
\end{equation}
For fermionic zero modes given by  $\Psi(x,y)=\psi(x) {k(y)}$ \cite{Melfo:2006hh}, current densities can be written as 
\begin{equation}\label{source}
J^c(x,y)=E^{\,\,\mu}_{\nu}\,\bar{\Psi}\,\gamma^{\nu}\,\Psi \,\delta^c_{\mu}
            =E^{\,\,\mu}_{\nu}\,k^2(y)j^{\nu}(x)\,\delta^c_{\mu},
\end{equation}
where $E^{\,\,\mu}_{\nu}$ is the vielbein and $j^{\nu}(x)$ is 
\begin{equation}
j^\nu(x)=\bar{\psi}(x)\gamma^\nu\psi(x).
\end{equation}
From (\ref{metric}) and (\ref{source}) we have  
\begin{equation}
J^c(x,y)=e^{-a(y)}\,k^2(y)j^{\mu}(x)\,\delta^c_{\mu}.
\end{equation}
If $\Psi(x,y)$ corresponds to a bound state, $k^2(y)$ should be peaked near the brane.

Now, for current densities of the form (\ref{source}), (\ref{generating}) reduces to
\begin{equation}
W[J]=W[0]\exp\left\{-\frac{1}{2}\int d^4p\,\tilde{j}^{\mu}(p)\left(\int dy\int dy' e^{3a(y)}k^2(y)\tilde{G}_{\mu\nu}(p,y,y')e^{3a(y')}k^2(y')\right)\tilde{j}^{\nu}(-p)\right\},
\end{equation}
where we have introduced four-dimensional Fourier transforms in Euclidean space. Thus, the quantity between parenthesis may be identified with the four dimensional gauge field propagator and we set
\begin{equation}\label{4-prop}
\tilde{G}^{(4)}_{\mu\nu}(p)= \int dy\int dy' e^{3a(y)}k^2(y)\tilde{G}_{\mu\nu}(p,y,y')e^{3a(y')}k^2(y').
\end{equation}
For $k^2(y)=\delta(y)$ we have
\begin{equation}\label{4-prop-lim}
\tilde{G}^{(4)}_{\mu\nu}(p)= e^{6a(0)}\tilde{G}_{\mu\nu}(p,0,0)
\end{equation}
and $\tilde{G}_{\mu\nu}(p,0,0)$ is essentially the four dimensional propagator.

As is well known, fermions cannot be gravitationally trapped in Randall-Sundrum branes.  However, if the   brane is  in fact a domain wall generated by the vacuum expectation value of a scalar field, as suggested in \cite{Rubakov:1983bb}, and fermions are coupled to this scalar field by a Yukawa term, 
confinement is possible, as shown in \cite{Bajc:1999mh} for the case of an (infinitely thin) kink. Now, a consistent kink profile can be obtained by requiring it to be a solution to the coupled Einstein-scalar field
system, with a suitable symmetry breaking potential \cite{DeWolfe:1999cp,Gremm:1999pj} and the thin
brane must be obtained as the thin wall limit of this solution \cite{Guerrero:2002ki}. However,  thin
wall geometries have distribution-valued curvatures whose singular
parts are proportional to a Dirac distribution supported on the
surface where the wall is localized. Strong conditions must be imposed  \cite{Geroch:1987qn} on a spacetime metric in order to ensure that its curvature tensors converge to the curvature tensors of the limit metric, and as a consequence domain wall solutions cannot be made infinitely thin while keeping
the asymptotic values of the scalar field fixed  \cite{Melfo:2006hh}.

For a domain wall solution of the coupled Einstein-scalar field equations, with an  asymptotically $AdS_5$ metric of the form (\ref{metric}), it is found that in order to have fermions on the wall, one can couple them to the scalar field $\phi$ with a Yukawa term of the form $\lambda \overline{\Psi}\Psi \phi$, and then get
\begin{equation}
{k_{\pm}(y)} = {\cal N} e^{- 2 a(y) \pm\lambda \int dy\, {\phi(y)}} \label{fmodes}. 
\end{equation}
For appropriate values of $\lambda$, the function ${k_{-}(y)}$ can in principle be normalized according to 
\begin{equation}
\int_{-\infty}^{\infty}dy\,e^{3a(y)}k^2_{-}(y)=1,
\end{equation}
as follows from the fermionic kinetic term of the action
\begin{equation}
\int d^4x\int dy \,\sqrt{g}\,\bar{\Psi}(x,y)i\Gamma^a\nabla_a\Psi(x,y)= \int dy \,e^{3a(y)}k^2(y)\int d^4x \,\bar{\psi}(x)i\gamma^{\mu}\partial_{\mu}\psi(x),
\end{equation}
and one chiral fermion mode gets confined whenever the wall's thickness is keep finite \cite{Melfo:2006hh}. Notice that for a $k^2_{-}(y)$ peaked near the brane, the quantity
\begin{equation}
e^{3a(y)}k^2_{-}(y)\propto e^{-a(y)-2\lambda\int dy\,\phi(y)}
\end{equation}
turns out to be even more peaked than $k^2_{-}(y)$ and (\ref{4-prop-lim}) may be a good approximation provided that $G_{\mu\nu}(p,y,y')$ is nonsingular and not too rapidly varying in the neighborhood of $y=0$ and $y'=0$.

In a flat spacetime with a brane at $y=0$, and assuming $k^2(y) = \delta(y)$, it is straightforward to integrate (\ref{g1})
\begin{equation}
G_1(p,y,y')=\frac{{\cal{ Q}}^2}{2p\,(1 + p/p^* )}u_1 (y_>) \,u _2( y_<)
\end{equation}
where $y_>=\mbox{max}\{y,y'\}$, $y_<=\mbox{min}\{y,y'\}$ and $u_1, u_2$ are given by
\begin{equation}
u_1(y)= \Theta(y){e^{-p y}} + \Theta(-y)\left[ \left(\frac{p}{p^*} + 1 \right)e^{-p y} -  \frac{p}{p^*} e^{p y}\right]  \; ,\qquad u_2(y) = u_1(-y),
\end{equation}
where $\Theta$ is the Heaviside distribution and $$p^*=2e^2/{\cal Q}^2.$$
It follows that
\begin{equation}\label{dgs}
G_1(p,0,0)=\frac{e^2}{p(p+p^*)},
\end{equation}
and $\tilde{G}_{\mu\nu}$ behaves for $p\gg p^*$ as the standard four-dimensional propagator, the well-known DGS result \cite{Dvali:2000rx}. 

\section{Brane-induced electromagnetism in the RS-2 escenario}

In order to study the effects of a warped geometry, let us consider the warped metric (\ref{metric}) with $a(y)=-\alpha|y|$, i.e., 
\begin{equation}
ds^2= e^{-2\alpha|y|}\ \eta_{\mu\nu}dx^{\mu}dx^{\nu}+ dy^2 \label{metric2}
\end{equation}
which can be regarded as the metric of the spacetime generated by an infinitely thin domain wall embedded in a $AdS_5$ spacetime with cosmological constant $\Lambda=-6\alpha^2$. Indeed, (\ref{metric2}) is the metric of the well known RS-2 scenario \cite{Randall:1999vf}. 

As a reference example, let us fix $k^2(y)=0$. In this case the kinetic term with support only on the brane is absent and the gauge field cannot be localized on the wall since, as is well known, vector fields cannot be gravitationally trapped in the RS scenario. For $k^2(y)=0$ we find 
\begin{equation}
G_1(p,y,y')=2\frac{{\cal{ Q}}^2}{\alpha}\frac{K_1(p/\alpha)I_1(p/\alpha)}{{\cal {F}}(p)}u_1 (y_>) \, u_2(y_<)
\label{g11}
\end{equation}
where $K_n$ and $I_n$ are the modified Bessel functions of order $n$,  $u_1, u_2$ are given by
\begin{eqnarray}
u_1(y)&=&\Theta(y)\frac{e^{\alpha y}}{K_1(p/\alpha)}K_1(e^{\alpha y}p/\alpha) + \Theta(-y)\left[{\cal{F}}(p)\frac{e^{-\alpha y}}{I_1(p/\alpha)}I_1(e^{-\alpha y}p/\alpha) + (1 -{\cal{F}}(p))\frac{e^{-\alpha y}}{K_1(p/\alpha)}K_1(e^{-\alpha y}p/\alpha)\right]  \; ,  \nonumber \\
 u_2(y)&=& u_1(-y)
\label{u11}
\end{eqnarray}
and $\cal{F}$ is found to be
\begin{equation}
 {\cal {F}}(p)= 2\frac{p}{\alpha}  K_0(p/\alpha) I_1(p/\alpha). 
\end{equation}
 
It follows that
\begin{equation}
G_1(p,0,0)\sim \frac{{\cal{ Q}}^2}{2}\frac{\alpha}{p^2\ln{(\alpha/p)}},\qquad p\ll\alpha,
\end{equation}
wich does not resemble the behaviour of a four-dimensional propagator in momentum space. On the other hand, we have 
\begin{equation}
G_1(p,0,0)\sim \frac{{\cal{ Q}}^2}{2}\frac{1}{p},\qquad p\gg\alpha,
\end{equation}
the five dimensional scaling in Minkowski spacetime.  This means that the effects of the curvature of the spacetime do not show up in the propagator at distances $r\ll\alpha^{-1}$.

Next, let us consider $k^2(y)=\delta(y)$. The propagator is again found to be given by (\ref{g11} -\ref{u11}), where now
\begin{equation}
 {\cal {F}}(p)= 2\frac{p}{\alpha}  K_0(p/\alpha) I_1(p/\alpha) \left( 1 + \frac{ K_1(p/\alpha)}{ K_0(p/\alpha)} \frac{p}{p^*}\right)
\end{equation}
and we find
\begin{equation}
G_1(p,0,0)= \frac{e^2}{p(p +p^*{K_0(p/\alpha)}/{K_1(p/\alpha))}}.
\end{equation}
For large momenta
\begin{equation}
G_1(p,0,0)\sim\frac{e^2}{p(p+p^*)},\qquad \alpha \ll p,
\end{equation}
which is the DGS result \cite{Dvali:2000rx}. As in the previous case, the effects of the curvature of the spacetime do not show up in the propagator at distances $r\ll\alpha^{-1}$. But at such short distances gravity ceases to be four dimensional for brane observers in the RS-2 scenario \cite{Randall:1999vf}. There is, however, another region in momentum space where this propagator behaves as the standard four-dimensional gauge propagator. We find  
\begin{equation}
G_1(p,0,0)\sim\frac{e^2}{p^2},\qquad \alpha\gg p \gg \frac{2e^2}{{\cal {Q}}^2}\frac{K_0(p/\alpha)}{K_1(p/\alpha)}\sim-\frac{2e^2}{{\cal Q}^2}\frac{\ln(p/\alpha)}{(p/\alpha)^{-1}},
\end{equation}
so that in the region 
\beq
\alpha\,e^{-\alpha/p^*} \ll p \ll \alpha
\label{rsbehavior}
\eeq
 four dimensional electromagnetism is recovered. Hence, brane observers in the RS-2 scenario that see four dimensional gravity will see also four dimensional electromagnetism for momenta in a region whose lower bound is gravitationally affected too. The additional dimension will show up in the electromagnetic interaction at very short distances $r \ll \alpha^{-1}$ and at large distances compared not with $r^*={p^*}^{-1}$ as in the DGS flat case but with  $\alpha^{-1}\exp{(\alpha r^*)}$.  
 
Effects in the electromagnetic propagator therefore set a limit $\alpha  \geq 10^{16}$ cm$^{-1}$ from the QED scale. 
 On the other hand, in \cite{Dvali:2000rx} it is estimated that 5 dimensional effects on the infrared behavior of the QED propagator would not be noticeable if the upper distance scale is even as low as solar-system size, or  $10^{15}$ cm. In view of (\ref{rsbehavior}), it would suffice to have
 \beq
\frac{Q^2}{e^2} \geq 10^{-14} cm
\eeq

\section{A thick domain wall in Minkowski space}

A domain wall solution in a five-dimensional Minkowski space, namely $a(y)=0$ in (\ref{metric}), is given by the scalar field 
\begin{eqnarray}
\phi(y)=\phi_0 \tanh(\alpha y/\delta) .
\end{eqnarray}
and the symmetry breaking potential 
\begin{eqnarray}
V(\phi)=\frac{1}{2}\ \beta (\phi_0^2-\phi^2)^2 , \label{V(phi)}
\end{eqnarray} 
where $ \phi_0 = \beta^{-1/2} \alpha/\delta$.  In this case
\beq
k^2(y) = \frac{\alpha}{ \sqrt{\pi}\delta}\frac{\Gamma\left(\lambda/\sqrt{\beta}+1/2\right)}{\Gamma\left(\lambda/\sqrt{\beta}\right)} \cosh(\alpha y/\delta)^{- 2 \lambda/\sqrt{\beta}}
\eeq
However, Eq. (\ref{g1}) for the Green function cannot be solved analytically. 

In order to exhibit the behavior of the solutions, we can choose the Yukawa coupling such that $\lambda = \sqrt{\beta}$. This particular choice is interesting since in the infinitely thin wall limit, $k^2(y)$ becomes $\delta(y)$, reproducing the DGS case. Now, exact solutions to the homogeneous version of  (\ref{g1}) can be found
\beq
 u_{1}(y)=   P_n^m\left(\tanh(\alpha y/\delta) \right),\qquad u_2(y)=u_1(-y), 
 \label{minksol}
\eeq
where $P_n^m(x)$ are the associated Legendre functions of degree $n$ and order $m$ with
\beq
n= - \frac{1}{2} \left( 1  \pm  \sqrt{1 - 4 \frac{p^2}{p^*} \frac{\delta}{\alpha} } \right),\qquad
m= \pm \frac{\delta}{\alpha}p
 \eeq
Boundary conditions demand that the negative sign be chosen in $m$, while both signs in $n$ give the same solutions. $G_1$ on the brane is then 
\beq
G_1(p,0,0) =  \frac{\delta e^2}{ 2 p^*\alpha}  \frac{\Gamma[ -(n+m)/2 ] \Gamma[ (1 + n-m)/2 ]}{\Gamma[ (1 - n-m)/2 ] \Gamma[ (2 + n-m)/2 ]}
\eeq
In the thin wall limit, $p \ll \alpha/\delta$, two cases may arise:
\begin{enumerate}
  \item $p^2 \gg p^* \alpha/\delta$. In this limit ordinary electromagnetism is never recovered, since
  \beq
  G_1(p,0,0) =  \frac{ e^2}{ p} \sqrt{\frac{\delta}{\alpha p^*}} 
  \eeq
  \item $p^2 \ll p^*  \alpha/\delta$. In this case we have
  \beq
  G_1(p,0,0) =  \frac{e^2}{  p (p + p^*)} \left[   1 + \frac{2 p^2 \delta \ln{2}}{p^* \alpha} + {\cal O} (p\,\delta/\alpha)^2 \right] 
  \eeq
  which gives the DGS result (\ref{dgs}) in the infinitely thin wall ($\delta\to 0$) limit.  
\end{enumerate}

It follows that in the region
\beq
p^* \ll p \ll \sqrt{p^*\alpha/\delta} 
\eeq
$G_{\mu\nu}$ behaves as the standard four-dimensional propagator. Hence, four dimensional electromagnetism is recovered in the region
\beq\label{flat-bounds}
 \sqrt{r^* r_c} \ll r \ll r^*
\eeq
where $r_c = \delta/\alpha$ is the wall's thickness and $r^* = {p^*}^{-1}$. The upper bound in (\ref{flat-bounds}) is the DGS electromagnetic critical radius while the lower one coincides with the result of \cite{Dubovsky:2001pe} and implies charge universality. However, we see now that the wall's thickness becomes related to the QED scale. Taking again the estimate of \cite{Dvali:2000rx}, namely   $r^* \geq 10^{15}$ cm, and the QED scale valid up to at least $10^{-16} $cm,  this gives upper bound to the wall's thickness of
\beq
r_c \ll 10^{-47}  {\rm cm} 
\eeq
This result seems to invalidate the DGS proposal to localize gauge fields on thick walls in a flat spacetime, at least in this particular case $\lambda=\sqrt{\beta}$.

\section{Conclusions}

The DGS mechanism of confinement of gauge fields on a brane is framed on two approximations: the brane is  infinitely thin, and the spacetime is flat. In this paper we have relaxed these approximations separately,  and investigated their effects on the critical scales at which electromagnetism is recovered. In order to do this, we have first presented a framework for calculating the electromagnetic propagator in a general thick brane in curved space. 

The inclusion of gravity was studied in a RS-2 scenario for an infinitely thin brane in a warped spacetime. As in the original DGS paper, fermions are assumed to be localized to the brane by a delta function. Both the ultraviolet and the infrared limits of the gauge field propagator are affected by the brane gravity, as could be expected by the inclusion of a new energy scale from the cosmological constant. This results  in a modification of the upper distance scale,   above which the five-dimensional effects on the electromagnetic propagator start to be important. This new scale  depends exponentially on the cosmological constant and the gauge couplings in 4 and 5 dimensions. The RS critical distance, below which 5-dimensional effects in the gravitational potential start to be measurable, now affects also the electromagnetic propagator. Since electromagnetism has been probed up to distances much smaller than the gravitational interaction, this sets a much more stringent bound on the brane's tension than the usual RS scenario.

Effects of the wall's thickness were then investigated by considering a thick wall solution to the Einstein-scalar field equations  where the bulk spacetime is Minkowski, which can be found by choosing a suitable scalar field potential. Fermions are then coupled to the brane by means of a Yukawa interaction with the scalar field, and the zero-modes can be calculated exactly. As could be expected, the wall's thickness plays a role in the ultraviolet limit. An analytical solution for the propagator can be found for a particular value of the ratio of Yukawa coupling to the scalar field self coupling. In this case, we recover the result of ref. \cite{Dubovsky:2001pe}, namely a critical scale below which the five dimensional effects become noticeable in the propagator, which is the product of the upper DGS critical scale  and the wall's thickness. This gives very stringent bounds on the wall's thickness, given that the DGS  scale has to be at least of the order of the solar system, and that electromagnetic interactions can be considered to be probed  to scales of around $10^{-16}$ cm. A wall would have to be many orders of magnitude thinner than Planck's scale in order to accommodate observations.   

Whether this conclusion holds for other values of the parameters should be determined by finding exact solutions for the propagator as a function of the Yukawa and self-coupling of the scalar field. Nevertheless, as the results of the warped spacetime case indicate, the inclusion of gravitation modifies both the ultraviolet and infrared behaviour of the effective four dimensional gauge field propagator in a brane world. Perhaps the confinement of gauge fields on a thick brane via the DGS mechanism, taking into account the gravitational interaction, imposes less stringent bounds.
 It would therefore be of interest to extend our work to the general case of thick walls in a warped spacetime, and general couplings.  
However the equations for the gauge field propagator become extremely involved, even for the most simple of thick brane solutions found in the literature, and cannot be solved analytically.  A numerical study is in order here. Work in this direction is in progress.

\section*{Acknowledgments}

This work was supported by CDCHT-UCLA under project 010-CT-2007.


\end{document}